\documentclass[sigconf,nonacm]{acmart}

\AtBeginDocument{%
  }

\usepackage{makecell}
\usepackage{subcaption}
\usepackage{tikz}
\usepackage{algorithm}
\usepackage{algpseudocode}

\usepackage{pifont}
\begin{document}

\title{Enhancing Document Retrieval for Curating N-ary Relations in Knowledge Bases}

\author{Xing David Wang}
\email{xing.david.wang@hu-berlin.de}
\orcid{0000-0001-8645-9354}
\affiliation{%
  \institution{Humboldt-Universität zu Berlin}
  \city{Berlin}
  \country{Germany}
}

\author{Ulf Leser}
\email{leser@informatik.hu-berlin.de}
\affiliation{%
  \institution{Humboldt-Universität zu Berlin}
  \city{Berlin}
  \country{Germany}}


\begin{abstract}
  Curation of biomedical knowledge bases (KBs) relies on extracting accurate multi-entity relational facts from the literature — a process that remains largely manual and expert-driven. An essential step in this workflow is retrieving documents that can support or complete partially observed n-ary relations. We present a neural retrieval model designed to assist KB curation by identifying documents that help fill in missing relation arguments and provide relevant contextual evidence.

  To reduce dependence on scarce gold-standard training data, we exploit existing KB records to construct weakly supervised training sets. Our approach introduces two key technical contributions: (i) a layered contrastive loss that enables learning from noisy and incomplete relational structures, and (ii) a balanced sampling strategy that generates high-quality negatives from diverse KB records. On two biomedical retrieval benchmarks, our approach achieves state-of-the-art performance, outperforming strong baselines in NDCG@10 by 5.7 and 3.7 percentage points, respectively.
\end{abstract}

\maketitle

\section{Introduction}

Curating information from scientific literature and its provisioning in the form of high-quality knowledge bases (KB) is an important task in biomedical research - one that can significantly benefit from the usage of efficient language processing algorithms~\cite{luo2022biored, li2016biocreative, lee2016bronco}. The curation process involves carefully identifying entities, their properties, and their relations to other entities described in a publication. For instance, knowledge bases in precision oncology (PO-KB) are filled with information regarding the relationships between genes, genomic variants, and treatments that may be applied to cancer therapy~\cite{lee2016bronco, lever_text-mining_2019}.

To reliably identify candidate papers that contain such information, KB curators typically query a repository of publications with the entities currently in their focus. A retriever ranks these documents based on their estimated relevancy to the curation task. The quality of this retriever is critical to the overall efficiency of the curation process, as it determines how many papers a curator must review on average to obtain a new KB relation~\cite{RE15}. 

State-of-the-art retrievers rely on neural models that require training. Three main techniques are used to obtain the necessary training data. First, high-quality gold standards can be obtained using manual annotation of query-document pairs, though this incurs high costs due to the necessary involvement of trained human experts. For instance, BioASQ~\cite{tsatsaronis2015overview} provides hand-crafted query-document pairs across a wide array of biomedical topics. Document references typically link to PubMed~\cite{NCBI_PubMed}, the primary repository for biomedical literature providing access to abstracts and metadata.

A second approach uses search engine logs, hypothesizing that documents that were clicked by users within the result to a query are also relevant to the query. The advantage of this technique is that query-document pairs emerge as a by-product of users searches and are thus free of charge; on the down-side, the resulting data is of more questionable quality, as clicking on a document does not automatically imply relevance, but rather is a first step to assess relevance. Furthermore, such click-log data typically is not publicly available due to privacy restrictions.
A system following this approach is MedCPT~\cite{jin_medcpt_2023} which was trained on query-document pairs extracted from PubMed user search logs.

A third approach, also followed by this work, is to use semi-supervised learning, in which records from the KB to be curated (or one from the same domain) are used to create positive training examples~\cite{bobic2012improving, abaho2023select}.
In contrast to prior work, we specifically exploit those KB relations that are already linked to external documents references providing grounding for the described relation and supervision signal to train a document retriever.
We treat the KB entities as queries and the linked documents as answers, thereby constructing (query, document) pairs for retriever training.

The textual evidence supporting a relation described by a KB record is often only found in the full text of the linked articles~\cite{griffith_civic_2017,chakravarty_oncokb_2017,uniprot2019uniprot}, which are typically not freely accessible\footnote{By license, automatic text mining is currently permitted for only ~7.5 million of the ~37 million articles indexed in PubMed, i.e., less than 25 percent.}.
The alternative is to use the more widely available abstracts; however, these may not fully contain the relevant entities and relational context. As a result, the supervision signal derived from such document references is inherently weak and noisy. Even in the presence of full texts, supervision is only available at the document level and not aligned to specific text spans.
To increase the quality of supervision signal in the available abstracts, we can filter out all query-document pairs where the entities of the query are not explicitly mentioned. For instance, when we turn the KB relation (\textit{DDR2, G253C, Dasatinib, 22328973}) for the (Gene, Variant, Treatment, PubMed-ID) entity types into a query-document pair, we first check whether each of the entities (\textit{DDR2, G253C, Dasatinib}) is contained in the linked abstract. If not, the pair is discarded from the training set.
In the PO-KB CIViC~\cite{griffith_civic_2017}, even when using extensive synonym lists for matching entities, only app. 1.000 of the 3.000 relationships between a gene, a genomic variant and a treatment can be matched completely to the corresponding abstracts of the linked PubMed IDs. This leads to a heavily reduced/highly noisy training data set compounding the challenge of weak supervision.

To address this issue, we introduce EDEL, a retrieval model specifically tailored to the KB curation task and designed to effectively deal with the noisy supervision signal. EDEL is based on a dense bi-encoder architecture~\cite{karpukhin_dense_2020,xiong_approximate_2020,hofstatter_efficiently_2021}, embedding queries and documents with a shared language model and computing relevance scores via similarity metrics like cosine similarity.

EDEL refines the standard dense retriever setup in two important aspects. First, it systematically increases training coverage by incorporating noisy positive examples — i.e., those query-document pairs where only a subset of entities is mentioned in the abstract. We assume that these partially matched examples still contain valuable training signals, and we account for their lower quality by using a layered contrastive loss that assigns them lower weights than fully matched examples.

Secondly, EDEL improves negative sampling by leveraging the structure of KBs: for each query-document pair derived from a KB, we collect documents from other KB records that share some but not all entities, treating them as hard negatives. This strategy is motivated by the observation that curators typically annotate exhaustively within a paper — so if two entities co-occur in a referenced paper but are not curated in the KB, the relation likely does not exist. We group these negatives by similarity and assign different margin penalties to encourage more discriminative learning.

To evaluate EDEL's effectiveness in supporting document retrieval during KB curation, we construct two benchmark data sets. The first dataset targets precision oncology (PO) and is based on the two KBs CIViC~\cite{griffith_civic_2017} and OncoKB~\cite{chakravarty_oncokb_2017}. Both contain curated data about treatment responses to specific genetic alterations. The second dataset models covers post-translational modifications (PTMs), which regulate protein function and enzyme activity. It is created from the important protein database UniProt KB~\cite{uniprot2019uniprot}. Both retrieval tasks aim to find evidence documents that help extend incomplete $(n-1)$-ary relations to full $n$-ary ones. 

We compare EDEL against three strong baselines: The two neural retrievers MedCPT~\cite{jin_medcpt_2023} and ColBERTv2~\cite{santhanam_colbertv2_2022}, and a bag-of-words BM25 retriever~\cite{robertson_probabilistic_2009}. Our results demonstrate the potential of our approach: EDEL outperforms the best competitor in both data sets by 5.7 and 3.7 percentage points (pp) in NDCG@10, and by 21.5 pp and 8.6 pp in Entity Recall@10. Ablation studies confirm that both model enhancements have positive effects.

The remainder of this paper is organized as follows: Section~\ref{sec:rel_work} reviews related work. Section~\ref{sec:datasets} introduces the KB curation task and our new datasets. Section~\ref{sec:methodology} details the EDEL model, including our layered contrastive loss and hard negative sampling strategy. Section~\ref{sec:experiments} presents the experimental setup and results and Section~\ref{sec:discussion} discusses our findings. We conclude with a summary and outlook in Section~\ref{sec:conclusion}.

\section{Related Work}
\label{sec:rel_work}

We begin by reviewing related work in information retrieval (IR), with a focus on models that use pre-computed vector embeddings for fast retrieval. We then provide context for our biomedical KB curation task by discussing methods that learn from knowledge bases, as well as retrieval models developed specifically for biomedical applications.

\textbf{Document retrieval.}
IR is a critical step for filtering large document collections and identifying relevant evidence for downstream tasks such as data curation~\cite{chen2017reading, izacard2023atlas}. IR pipelines typically use a lightweight first-stage retriever, followed by a more computationally expensive second-stage re-ranker for more accurate document ranking~\cite{nogueira2019passage}.

BM25~\cite{robertson_probabilistic_2009}, a sparse vector space model based on lexical features, remains a strong baseline for first-stage retrieval~\cite{thakur2021beir, rajapakse2024negative}. In contrast, dense retrievers use neural networks to embed queries and documents into high-dimensional vector spaces, enabling semantic similarity-based retrieval~\cite{karpukhin_dense_2020, izacard2021unsupervised, hofstatter_efficiently_2021, jin_medcpt_2023}. \citet{karpukhin_dense_2020} and \citet{thakur2021beir} showed that dense retrievers alone can be competitive with traditional two-stage pipelines. These models are typically trained with supervised query-document pairs and relevance labels. Unsupervised approaches such as Contriever~\cite{izacard2021unsupervised} use data augmentation to construct training signals from unlabeled documents.

Effective negative sampling is crucial for both supervised and unsupervised retrievers. DPR~\cite{karpukhin_dense_2020} demonstrates that combining in-batch negatives with BM25 negatives improves performance over random sampling. ANCE~\cite{xiong_approximate_2020} adopts a dynamic strategy, using top-ranked negatives from earlier model iterations to refine training. TAS-B~\cite{hofstatter_efficiently_2021} uses knowledge distillation from a powerful teacher re-ranker and samples diverse negatives uniformly from the whole margin score distribution. It groups related queries into batches to enhance in-batch negative sampling. In contrast to TAS-B, EDEL can derive a set of balanced negatives from a structured KB without requiring access to margin scores from a teacher model.

ColBERTv2~\cite{santhanam_colbertv2_2022} is a neural retriever that uses token-level rather than document-level embeddings. It employs a late interaction mechanism to combine token embeddings at query time. ColBERTv2 demonstrates strong generalization in zero-shot settings on benchmarks such as BEIR~\cite{thakur2021beir} and is used as a teacher model for retrievers like TAS-B.

\textbf{Biomedical IR.}
Biomedical research relies heavily on high-quality IR systems. BioASQ~\cite{tsatsaronis2015overview} is a long-running shared task series that promotes IR and question answering in the biomedical domain. TREC tracks such as TREC COVID~\cite{voorhees2021trec}, TREC Precision Medicine~\cite{roberts_overview_2019}, and TREC Clinical Trials~\cite{roberts2022overview} have targeted domain-specific retrieval challenges.

Several IR models have been developed specifically for biomedical tasks. VIST~\cite{seva_vist_2019} combines keyword retrieval with document classification for precision oncology. MedCPT~\cite{jin_medcpt_2023} is a dense retriever trained on over seven million queries and 255 million query-document pairs from PubMed search logs. It achieved strong results on biomedical benchmarks in BEIR, outperforming other retrievers. We compare EDEL to both the pre-trained MedCPT model and versions fine-tuned on each of the two KB datasets.

\textbf{Semi-supervised learning using knowledge bases.}
Using knowledge bases for semi-supervised learning in question answering and IR has been a popular approach since more than two decades~\cite{mintz2009distant, surdeanu2012multi, weber2020pedl}. Approaches like \citet{ding2024enhancing} and \citet{nangi_dense_2023} leverage knowledge bases to tune dense retrieval models for question answering over knowledge graphs and approaches like \citet{abaho2023select} tune them to extend knowledge graphs directly. However, this focus on graph links disregards the need for supporting external evidence documents, which is crucial for high-quality data curation~\cite{cerami2010pathway, griffith_civic_2017, chakravarty_oncokb_2017}.

\textbf{Margin losses.}
Various loss functions have been proposed to learn effective query-document representations. Pairwise losses enforce a margin of at least $m$ between positive and negative examples~\cite{hofstatter_efficiently_2021}, while adaptive versions dynamically adjust $m$ during training based on the current positive and negative example~\cite{ha2021deep}. Pointwise losses~\cite{hadsell2006dimensionality} directly enforce a fixed margin $m$ between the query representations and the given negative/positive example. In EDEL, we extend such pointwise losses by introducing adaptive, instance-specific margin values for each example instead of relying on fixed margins.

\section{Datasets and Task Definition}
\label{sec:datasets}

Reliable extraction of $n$-ary relation structures has garnered an increasing interest in the recent years and is particularly relevant in the biomedical domain~\cite{peng2017cross,luo2022biored}. However, existing benchmark datasets are not designed to support the retrieval of such structured relations, but instead focus more on general question answering like BioASQ~\cite{tsatsaronis2015overview}. Datasets developed for relation extraction, like N-ary~\cite{peng2017cross} and BioRED~\cite{luo2022biored}, are also limited in size (e.g., $\sim$600 abstracts) due to the high annotation cost, making them impractical for training large-scale retrieval models.

To address this gap, we introduce two new biomedical retrieval datasets: one for Precision Oncology (PO), focusing on ternary relations, and one for Post-Translational Modifications (PTM), involving more complex relations with four entities.

\subsection{Dataset: Precision Oncology (PO)}

Precision oncology operates on the premise that for each pathogenic genetic alteration in a patient’s cancer genome, there may exist a targeted treatment. As the number of identified alterations and available therapies continues to grow, clinicians in molecular tumor boards often spend considerable time reviewing literature and knowledge bases for each patient case~\cite{rieke2022feasibility}. However, manual curation is time-consuming, leading to incomplete KBs~\cite{rieke2022feasibility}. Our PO retrieval task targets the first step of the curation process: identifying candidate documents that describe treatments for specific gene-variant combinations.

We use the CIViC~\cite{griffith_civic_2017} and OncoKB~\cite{chakravarty_oncokb_2017} knowledge bases to construct a new benchmark for retrieval in the KB curation setting. Both KBs provide expert-curated annotations on cancer genes, variants, and when available, associated treatments\footnote{We use the CIViC snapshot from 01/11/22 (\url{https://civicdb.org/downloads/01-Nov-2022/01-Nov-2022-ClinicalEvidenceSummaries.tsv}) and the OncoKB snapshot from 07/06/23 (available on request at \url{https://www.oncokb.org/api-access}).}. For this dataset, we focus on three entity types: Gene, Variant, and Treatment, where Treatment is the target entity. Although CIViC has previously been used in benchmarks for relation extraction~\cite{peng2017cross,lever_text-mining_2019}, it has not yet been adapted for IR.

\subsection{Dataset: Post-Translational Modification (PTM)}

A post-translational modification (PTM) is a modification to the structure and composition of a protein after it was created by translation from mRNA. PTMs are important processes for modulating the function of a protein. For instance, protein functions can be regulated by adding functional groups to individual amino acids (called residue) in the protein sequence, e.g., phosphoryl groups during a protein phosphorylation (PTM type), which  work like switches between different potential functions~\cite{cohen2002origins}. Other PTM types include acetylation and methylation. PTMs are often driven by other helper proteins called catalysts.

To support retrieval in the context of PTM curation, we constructed a second dataset using the UniProtKB knowledge base~\cite{uniprot2019uniprot}, which contains structured data about proteins, their sequences, PTMs, and protein-protein interactions\footnote{We use the snapshot from 27/03/24: \url{https://ftp.uniprot.org/pub/databases/uniprot/current_release/knowledgebase/complete/uniprot_sprot.dat.gz}.}. The curated relations in this dataset involve four entity types: Substrate protein, PTM type, Residue, and Catalyst.

\subsection{From KB records to query-document pairs}
\label{ssec:query_generation}

We leverage structured KB records to construct positive query-document pairs $(q, d)$ for training an IR model. The task is to retrieve supporting evidence documents for extending partially observed $n$-ary relations ($n > 2$). Given a relation schema $r = (e_1, e_2, ..., e_{n-1}, \mathbf{e_n}, d)$ with $e_1, ..., \mathbf{e_n}$ as entity instances of different types and $d$ as a document reference, we aim to retrieve documents that complete a partial $(n{-}1)$-ary relation $(e_1, ..., e_{n-1})$ by adding a target entity $\mathbf{e_n}$. We refer to the first $n{-}1$ entities as the query entities $r_q$ and to the final one as the answer entity $r_{ans}$.

We emphasize that our scope here is document retrieval only: the model retrieves candidate documents $d$, but does not extract the answer entities $r_{ans}$ from them. Since standard IR metrics such as MAP and nDCG evaluate retrieval based only on a predefined set of gold documents (and may penalize retrieval of unannotated but valid documents), we additionally introduce a metric called EntityRecall, which better reflects downstream extraction performance (see Section~\ref{ssec:eval_measure}).

EDEL converts KB records into natural language queries using predefined templates that define the target entity type ($\mathbf{e_n}$) and insert placeholders for the query entities ($e_1, ..., e_{n-1}$). For example, in the PO dataset, we use the template \textit{Treatment for gene $e_1$ and variant $e_2$?} to create queries that retrieve candidate documents for gene-variant pairs. In this case, Gene and Variant are the query entities, while Treatment is the answer entity.
Each such created query $q$ is then linked to the referenced document $d$ in the KB to create a positive query-document pair\footnote{We discard incomplete KB records that lack answer entities $\mathbf{e_n}$. Note that multiple records can share the same query entities and thus result in duplicate queries.} $(q,d)$.

A similar example in the PTM dataset is: \textit{Catalysts for the phosphorylation of AKT1 at serine 473?}, where \textit{phosphorylation} is the PTM type, \textit{AKT1} is the Substrate protein, and \textit{serine 473} is the Residue\footnote{Since gene and protein names are ambiguous, multiple representations (e.g., gene symbols, full names, synonyms) may be used. Based on preliminary experiments, we use NCBI’s official gene symbol in PO queries, and both the UniProt symbol and full name in PTM queries.}. The answer entity type is Catalyst.

\subsection{Dataset Statistics}

Table~\ref{tab:dataset_statistics} reports the number of unique gene/protein entities, queries, query-document pairs, and entity tuples ($e_1, ..., e_{n-1},\mathbf{e_n}$) encountered in each dataset. Due to the small number of unique queries in OncoKB, we merge it with CIViC and treat both as a unified dataset called PO for training and evaluation.
We count approximately three variants per gene and four relevant documents per gene-variant query in the PO dataset, yielding ~3,300 positive query-document pairs. In the PTM dataset, we generate one query per protein on average and count two supporting documents per query, resulting in ~4,500 positive query-document pairs.

To prevent data leakage, we ensure that all queries and query-document pairs involving the same gene or protein are assigned to the same split. We divide each dataset into train, development, and test sets. The final test sets contain 133 queries for PO and 480 queries for PTM.

\begin{table}[htbp]
  \centering
  \caption{Number of unique gene/protein entities, queries, positive query-document pairs and entity tuples extracted from the two precision oncology (PO) KBs CIViC, OncoKB and the UniProt KB for post-translational modifications (PTM).}
    \resizebox{\linewidth}{!}{%
    \begin{tabular}{l|rrr|r}
    \toprule          & \multicolumn{3}{c|}{\textbf{PO}} & \multicolumn{1}{c}{\textbf{PTM}} \\
          & \multicolumn{1}{l}{\textbf{CIViC}} & \multicolumn{1}{l}{\textbf{OncoKB}} & \multicolumn{1}{l|}{\textbf{All}} & \textbf{Uniprot} \\
    \midrule
    Gene/Protein entities $e_1$ &                  333  &                  158  &                  395  &              2,002  \\
    Queries $r_q, q$ &                  939  &                  164  &              1,064  &              2,455  \\
    Positive pairs $(q,d)$ &              2,126  &              1,187  &              3,261  &              4,492  \\
    \midrule
    Entity tuples $(e_1, ..., \mathbf{e_n})$ & 2,262 & 458 & 2,666 & 2,861 \\
    \bottomrule
    \end{tabular}%
    }
  \label{tab:dataset_statistics}%
\end{table}%

\section{Methodology}
\label{sec:methodology}

\begin{figure*}[ht]
\centering
\begin{tikzpicture}[node distance=1.5cm, auto]

\node[minimum width=3cm, minimum height=5cm, baseline] (table) {
    \begingroup
    \scriptsize
    \setlength{\tabcolsep}{2pt}
    \begin{tabular}{cccc}
        \toprule
        \multicolumn{4}{l}{Knowledge base with records}  \\
        \midrule
        Gene & Variant & Treatment & PubMed ID \\ \midrule
        \textbf{SMO} & \textbf{L412F} & Vismodegib & 26822128 \\
        \textbf{SMO} & \textbf{L412F} & Vismodegib & 25759020 \\
        \textbf{SMO} & D473H & Saridegib & 22550175 \\
        BRAF & L597R & Trametinib & 22798288 \\
        ... & ... & ... & ... \\ \bottomrule
    \end{tabular}
    \endgroup
};

\node[draw, rectangle, rounded corners, fill=blue!20, text width=5.3cm, align=left, inner xsep=2mm, minimum height=5cm, baseline, right of=table, xshift=5cm] (textbox) {
    \scriptsize
    \textbf{\textcolor{blue}{Query: Treatment for gene SMO and variant L412F?}} \\
    \textbf{Positive sample classes} \\
    1. Gene, Variant and Treatment all matching, $\mu =0.0$ \\
    {\tiny - Doc 26822128: [...] The \textbf{p.L412F} mutation was found experimentally to result in increased \textbf{SMO} transactivating activity, and the patient responded to \textbf{vismodegib} therapy. [...]} \\
    2. Only Gene matching, $\mu <1.0$ \\
    {\tiny - Doc 25759020: [...] we show that both classes of \textbf{SMO} variants respond to aPKC-$\iota$/$\lambda$ or GLI2 inhibitors that operate downstream of \textbf{SMO} [...] } \\
    \textbf{Negative sample classes} \\
    1. Same Gene, Other Variant (matching), Any Treatment, $\mu >0.6$ \\
    {\tiny - Doc 22550175: [...] \textbf{saridegib} was found to be active in cells with the \textbf{D473H} point mutation that rendered them resistant to another \textbf{Smo} inhibitor, GDC-0449 [...]} \\
    2. Other Gene, Any Treatment, $\mu >0.8$ \\
    {\tiny - Doc 22798288: [...] This study shows that cells harboring BRAF(L597R) mutants are sensitive to MEK inhibitor treatment [...]} \\
};

\node[minimum width=3cm, minimum height=5cm, baseline, right of=textbox, xshift=5.5cm] (figure) {
    \includegraphics[width=5cm]{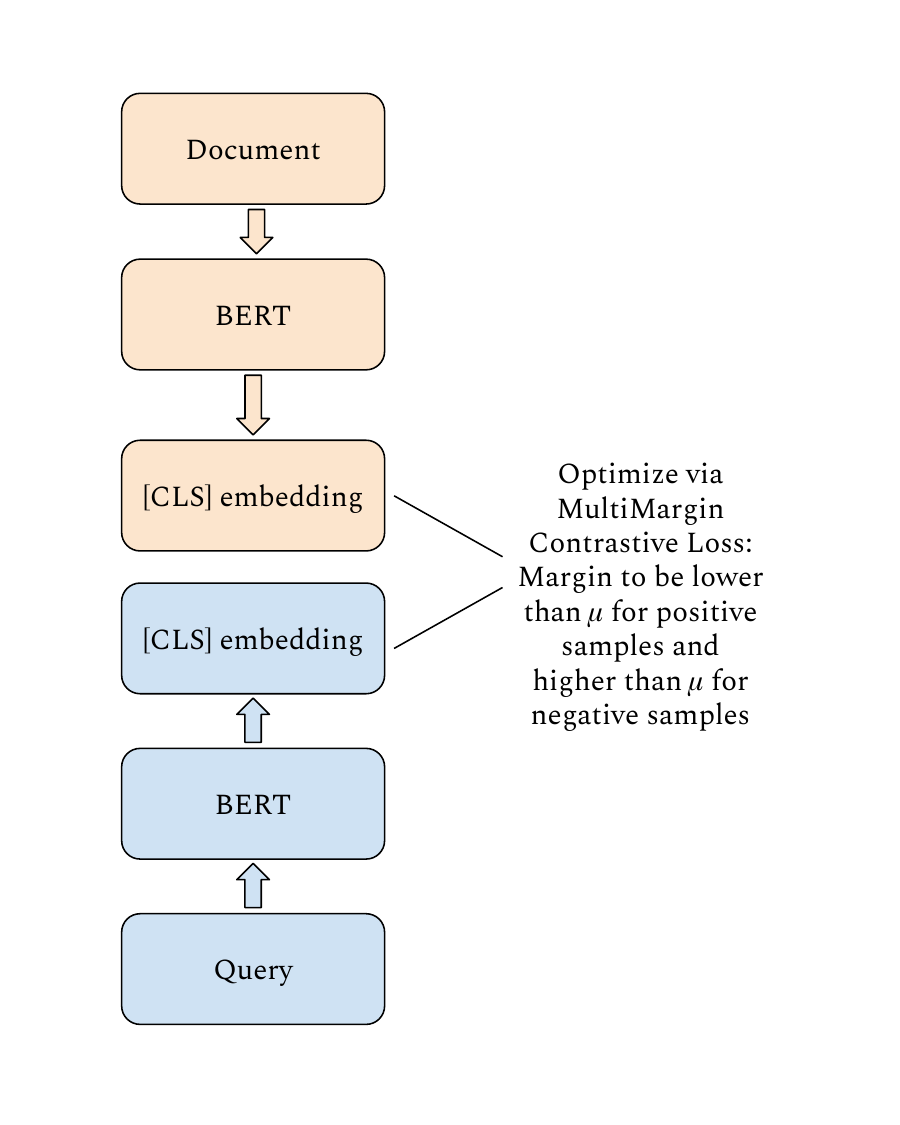}
};

\draw[->, line width=1mm, shorten >=5pt, shorten <=5pt, draw=blue!50!cyan] (table.east) -- (textbox.west);
\draw[->, line width=1mm, shorten >=5pt, shorten <=5pt, draw=blue!50!cyan] (textbox.east) -- (figure.west);

\end{tikzpicture}
\caption{Overview of the EDEL framework. Left: Negative samples are grouped into predefined classes and assigned margin values $\mu$ based on their overlap with query/answer entities. Right: Query and document embeddings are optimized using a MultiMargin contrastive loss.}
\label{fig:flowchart}
\end{figure*}
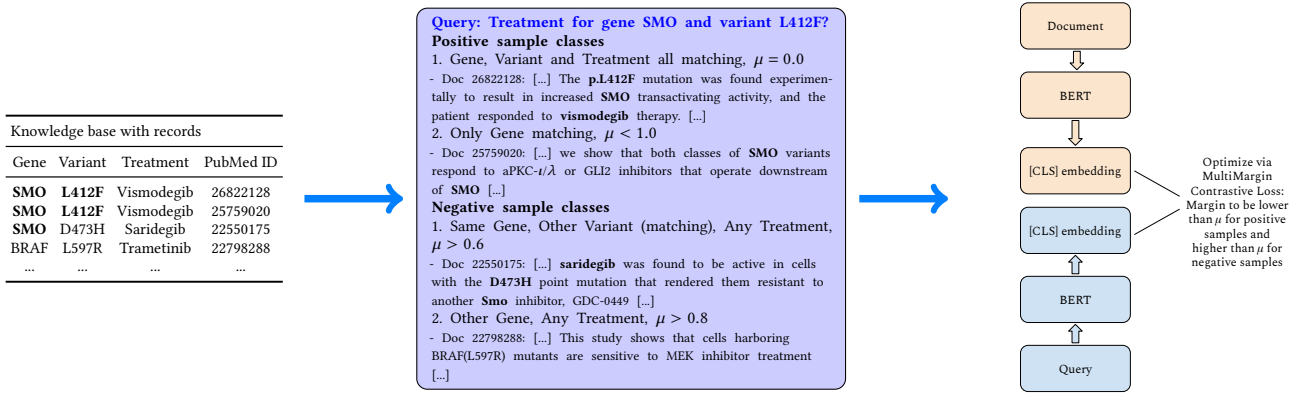

We present EDEL, a new method to efficiently leverage structured knowledge bases (KBs) for training dense retrievers models. EDEL addresses the challenging task of retrieving evidence documents that support the curation of $n$-ary relations. It introduces two key innovations: Given positive query-document associations from a KB, (i) EDEL uses a layered, contrastive loss function to handle noisy positives with different relevance levels and (ii) EDEL carefully balances negative sample selection by drawing from both hard and easy negatives. An overview of our workflow is shown in Figure~\ref{fig:flowchart}. 

\subsection{Noisy positives and layered margin loss}
\label{ssec:noisy_positives}

To capture the relevance between queries and documents, we use cosine distance as the scoring function, defined as:
\begin{align}
 \text{dist}\big(\vec{v}(q), \vec{v}(d) \big) = 1 - \frac{\vec{v}(q) \cdot \vec{v}(q)}{|\vec{v}(q)|~|\vec{v}(q)|},
\end{align}
where $\vec{v}(q)$ and $\vec{v}(d)$ are the [CLS] token embeddings of query $q$ and document $d$, $\cdot$ denotes the scalar product and $|\vec{v}(q)|$ the vector norm. Cosine distance ranges from 0 (identical angles, i.e., positive document) to 2 (opposite angles, i.e., negative document).

A major challenge in biomedical IR is that most KBs reference full-text documents, but only ~25 percent of the full-text documents are openly accessible. Abstracts are always available, but often do not mention all KB-linked entities, leading to noisy positives, i.e., documents containing only a subset of all the query and answer entities. In our datasets, noisy positives account for ~80 percent of examples in PO and ~87.5 percent in PTM. 

Rather than discarding these instances, we incorporate them using a layered contrastive loss, which assigns different margin values to positive examples depending on how completely the abstract covers all the KB entities. This approach reflects a graded notion of relevance rather than a binary one.

We define the textual match function $g(e, d) \in \{0, 1\}$ that returns 1 if entity $e$ (or any of its synonyms) is mentioned in document $d$, and 0 otherwise. For a given KB record $(e_1, ..., e_n, d)$, we evaluate $g$ for each entity type and assign an overall margin $\mu$ based on the number of matching entities. If all entities evaluate to 1 under $g$, the margin $\mu = 0$ is assigned to it (a completely matching positive). If one or more entities evaluate to 0, we assign a margin higher than 0, dependent on the number and the type of the omitted entities. Examples with the same evaluations under $g$ form groups of so-called positive margin classes.

We introduce four noisy positive classes for both the PO and PTM datasets in addition to the complete positives mentioned above, each with a distinct margin value ranging between 0 and 2, thus balancing the total amount of classes between model expressivity and complexity.
For instance, consider the KB record \textit{(SMO, L412F, Vismodegib, PubMed ID 26822128)} from Figure~\ref{fig:flowchart}. If the treatment \textit{Vismodegib} is not mentioned in the abstract but all other entities are, then we consider it a noisy positive. We assign a margin value of 0.2 higher than 0 to the corresponding query-document pair. If neither the variant \textit{L412F} nor the treatment \textit{Vismodegib} are mentioned in the abstract, then we consider it a noisy positive again albeit less relevant than the one before and assign it an even higher margin value such as 0.6. Further details about all the margin classes and their specific margin values are provided in Table~\ref{tab:margin_classes_stats} and Appendix~\ref{app:ssec:margin_classes}.

\begin{table}[htbp]
\setlength\tabcolsep{4pt} 
\caption{Margin classes, their margin values $\mu$ and number of positive and negative examples for each of the two datasets PO and PTM.}
\resizebox{0.8\linewidth}{!}{%
  \begin{tabular}{lr|rr|rr}
    \toprule
    & Dataset & \multicolumn{2}{r|}{PO} & \multicolumn{2}{r}{PTM} \\
    Class & $\mu $ & Pos & Neg & Pos & Neg \\
    \midrule
    1  & 0.0 & 827 & 0 & 563 & 0 \\
    2  & 0.2 & 2,065 & 16,775 & 1,179 & 2,015 \\
    3  & 0.6 & 218 & 9,912 & 1,405 & 1,260 \\
    4  & 0.8 & 0 & 149,083 & 1,294 & 45,736 \\
    5  & 1.0 & 1,117 & 176,079 & 51 & 43,473 \\
    6  & 1.2 & 200 & 0 & 0 & 37,980 \\
    \midrule
    \multicolumn{2}{l|}{Total} & 4,427 & 351,849 & 4,492 & 130,464 \\ 
    \bottomrule
  \end{tabular}%
  \label{tab:margin_classes_stats}
}%
\end{table}

To handle these distinct margin values, we extend the classic contrastive loss function~\cite{hadsell2006dimensionality} to our MultiMargin variant:  
\begin{align}
    & \ell_{MultiMargin} (q, d) = \\
    &  l(d) \cdot \text{max}\{0, [\text{dist} \big(\vec{v}(q), \vec{v}(d)\big) - \arccos(1- \mu(d)]^2) \} \nonumber \\
    & +  (1 - l(d)) \cdot \text{max}\{0, [\arccos(1-\mu(d)) - \text{dist} \big(\vec{v}(q), \vec{v}(d)]^2\}. \nonumber
\end{align}
\label{eq:loss}
where the label $l(d)=1$ for positive samples $d$, $l(d)=0$ for negative samples, and $\mu(d)$ being the flexible margin value dependent on the specific class of the associated document.

This loss encourages closer alignment between queries and relevant documents (with smaller margins) and separation from less relevant or negative ones (with larger margins). Compared to margin-based distillation methods like TAS-B~\cite{hofstatter_efficiently_2021}, EDEL's definition of margin values can be seen as a kind of soft labeling mechanism a priori, without relying on a teacher model to provide margin scores.

\subsection{Hard negatives sampling}
\label{ssec:hard_negatives}

Sampling hard negatives is essential for improving dense retriever performance~\cite{karpukhin_dense_2020, xiong_approximate_2020, hofstatter_efficiently_2021}. For each positive pair, we select informative negative documents to train the model to discriminate between relevant and non-relevant documents.

EDEL leverages the structured nature of KB records to identify high-quality negatives for a given positive query-document 
pair $(q,d)$: Given the KB record $r=(e_1, ...,$ $e_{n-1},$ $\mathbf{e_n}, d)$ associated to $(q,d)$, we consider all other records $k$ from the KB with (i) either different query entities or (ii) no answer entities as potential negatives. For example, another document $d_k$ mentioning the same variant \textit{SMO L412F} as the gold document $d$ but no treatment is likely irrelevant for treatment retrieval and thus serves as a negative.
We generalize this idea by introducing the KB matching function $f(e_1, e_2)$ which compares two entities $e_1$ and $e_2$ of the same type from two different KB records: For query entities, $f = 1$ if both entities coincide, otherwise $f = 0$. For answer entities, we relax the condition and set $f = 1$ if any entity is filled in the corresponding answer slots $e_1$ and $e_2$, as there can exist multiple, valid answer entities extending the $(n-1)$-ary relation.
Then, we can not only use $f$ to determine similarity between individual entity types but also to distinguish positive and negative sample documents $d_k$ from the other records $k$ given the positive record~$r$:  If $f = 1$ for all entity types, then $d_k$ is another positive sample.
If $f = 0$ for at least one entity type, then $d_k$ is a negative sample.\footnote{False negatives may occur due to incomplete KB curation. However, we assume that documents cited in a KB have been fully curated, and thus missing entity associations are unlikely.}

\textbf{Defining negative sample classes.}
We further distinguish negatives into different sample classes based on their evaluation of the function $f$. We sample separately from each of those classes ensuring a diverse set of negatives to be encountered during training. For example, all negatives that do not share any query entity with the positive KB record $r$ and do not contain any answer ($f=0$ for all entities) form the class of easy negatives. Negatives that share all the same query entities as the positive record ($f=1$) but do not contain any answer ($f=0$) form a hard negative class; those that share no query entities but contain at least one answer form another one. Classes that share more entities with the given positive record are treated as potentially harder examples as they are more likely to be confused with positive examples during training. We assign individual margin values $\mu$ up to 1.2 (a cosine distance of 1.0 already implies orthogonality, i.e., unrelatedness) for each class to model varying relevance levels and tune those values via grid search.

For ease of notation, we further group classes with same margin values together. We sample up to a maximum of $m=50$ negatives per class for each positive example. Our final setup includes four negative classes for PO and five for PTM, ranging from hard to easy samples. Details and class-wise margin values are shown in Table~\ref{tab:margin_classes_stats} and Appendix~\ref{app:ssec:negatives}.

\section{Experiments}
\label{sec:experiments}

\subsection{EDEL model architectures}
We train EDEL using a bi-encoder architecture with one shared encoder for both query and document embeddings. The encoder is initialized from the BioLinkBERT-base checkpoint~\cite{yasunaga-etal-2022-linkbert} containing 110 M parameters. 
We fine-tune separate retriever models on each KB task, one on the PO dataset, the other one on the PTM dataset.

To evaluate the contribution of each individual EDEL component, we compare the full EDEL model to model ablations, where we leave out increasingly more components: The full EDEL model employs (i) hard negatives (HardNeg) and (ii) noisy positives (NoisyPos) together with (iii) the layered, contrastive loss function (MultiMargin). Correspondingly, the EDEL (-MultiMargin) model only uses binary relevance margins with $\mu=0$ for all positive samples, and $\mu=0.8$ for all negative examples\footnote{We also remove hard negatives that closely resembles noisy positives (class 2 in Table~\ref{tab:margin_classes_stats}) so there is a clear decision boundary between positive and negative examples.}. Additionally, EDEL (-MultiMargin,-NoisyPos) removes noisy positives and their corresponding negative samples whereas EDEL (-MultiMargin,-NoisyPos,-HardNeg) replaces the whole hard negative sampling approach by random negatives.
More details on model training and how to reproduce our results can be found in the Appendix~\ref{app:sec:training}.

\subsection{Baselines}
We first compare EDEL to baselines in a zero-shot setting to show that the KB retrieval task is not trivial and benefits from domain-specific fine-tuning.
We evaluate the performance of three strong baselines BM25~\cite{robertson_probabilistic_2009}, a classic retriever model based on bag-of-words, ColBERTv2~\cite{santhanam_colbertv2_2022}, a general purpose dense retriever using late interaction, and MedCPT~\cite{jin_medcpt_2023}, a dense retriever specifically trained on PubMed, in a zero-shot setting.

Secondly, we compare EDEL to other fine-tuning approaches to show that our proposed training method indeed provides tangible benefits for more precise document retrieval. For that, we compare EDEL to a variant of the MedCPT retriever that we further fine-tune on the KB query-document pairs\footnote{We continue MedCPT training by further fine-tuning the retriever checkpoint using the same set of (complete) positive examples and in-batch negatives with a batch size of $n$=32. For fine-tuning, we use the original training script provided by the authors under \url{https://github.com/ncbi/MedCPT/blob/main/retriever/main.py}.}.

\subsection{Evaluation metrics}
\label{ssec:eval_measure}

Following popular retrieval benchmarks like BEIR~\cite{thakur2021beir} and MTEB \cite{muennighoff2022mteb}, we report the NDCG (non-discounted cumulative gain) and MAP (mean average precision) at $k=10$ and $k=50$. Both NDCG and MAP quantify how well our retrieval systems can identify the referenced gold document in the corresponding KB record for a given query. We put the main focus on the NDCG metric which puts more emphasis that relevant documents get assigned a high ranking, as this is the most important property to achieve in curation. 

However, neither NDCG nor MAP can assess how well our systems are able to actually retrieve abstracts that directly contain relevant entity information in them as the gold documents their evaluations are based on consist of both noisy positive and complete positive documents. As a dedicated metric to reflect this important property of a retrieval model targeting KB curation, we introduce the Entity Recall metric and evaluate it at $k=10$ and $k=50$. EntityRecall measures how many answer entities $r_{ans}$ from the KB record can be actually found in the top-k retrieved documents. We count an answer entity as successfully found if (i) there is at least one mention of the answer or one of its synonyms in any of the top-k retrieved documents and (ii) there is at least one mention of the corresponding query gene/protein, i.e., indicating there is likely a relation described in the document involving both query and answer entities.
Entity Recall thus accounts for potentially missed relevant documents during initial curation by human experts as it only evaluates entity mentions and does not evaluate on a fixed set of gold documents.
It also serves as a proxy to evaluate retriever performance in potential downstream tasks like question answering and Retrieval-Augmented Generation contexts~\cite{ram2023context, ghossein2024iclerb}.

\begin{table*}[htbp]
  \centering
  \caption{Performances on the datasets Precision Oncology and Post-translational modifications measured by NDCG, MAP and Entity Recall @10 and @50 in \%. n denotes the number of queries. Best score in each column is marked in bold.} 
    \begin{tabular}{l|rr|rr|rr|rr|rr|rr}
    \toprule
    \textbf{Dataset} ($\rightarrow$) & \multicolumn{6}{c|}{\textbf{Precision Oncology}} & \multicolumn{6}{c}{\textbf{Post-translational Modifications}}  \\
    & \multicolumn{2}{c|}{NDCG} & \multicolumn{2}{c|}{MAP} & \multicolumn{2}{c|}{Entity Recall} & \multicolumn{2}{c|}{NDCG} & \multicolumn{2}{c|}{MAP} & \multicolumn{2}{c}{Entity Recall} \\
    \textbf{Model Name} ($\downarrow$) & \multicolumn{1}{c}{@10} & \multicolumn{1}{c|}{@50} & \multicolumn{1}{c}{@10} & \multicolumn{1}{c|}{@50} & \multicolumn{1}{c}{@10} & \multicolumn{1}{c|}{@50} & \multicolumn{1}{c}{@10} & \multicolumn{1}{c|}{@50} & \multicolumn{1}{c}{@10} & \multicolumn{1}{c|}{@50} & \multicolumn{1}{c}{@10} & \multicolumn{1}{c}{@50}   \\ \midrule
    \multicolumn{13}{c}{\textbf{Zero-shot models}} \\ \midrule
    BM25  & 11.17 & 12.37 & 9.07 & 9.29 & 21.21 & 43.56 & 20.50 & 23.70 & 16.70 & 17.47 & 42.11 & 55.73 \\
    ColBERTv2 & 10.95 & 12.46 & 9.03 & 9.32 & 31.82 & 53.79 & 18.59 & 21.60 & 14.78 & 15.49 & 39.96 & 50.54 \\
    MedCPT & 9.15 & 11.18 & 7.57 & 7.89 & 27.27 & 48.48 & 13.10 & 17.30 & 10.07 & 11.05 & 35.13 & 48.39 \\ \midrule
    \multicolumn{13}{c}{\textbf{Fine-tuned models with KB entries}} \\ \midrule
    MedCPT & 12.91 & 15.46 & 10.87 & 11.34 & 28.41 & 53.79 & 19.64 & 23.23 & 15.73 & 16.54 & 40.14 & 57.53 \\ 
    EDEL  & \textbf{18.62} & \textbf{23.66} & \textbf{14.07} & \textbf{15.24} & \textbf{53.41} & \textbf{73.48} & \textbf{24.24} & \textbf{26.47} & \textbf{19.73} & \textbf{20.21} & \textbf{50.72} & \textbf{61.65} \\
    \bottomrule
    \end{tabular}%
  \label{tab:20250226_results_metrics}%
\end{table*}%

\subsection{Results}
\label{ssec:results}

Table~\ref{tab:20250226_results_metrics} presents the results of our fine-tuned EDEL model and the baseline retrievers, evaluated in both zero-shot and fine-tuned settings on the two KB retrieval tasks.

\textbf{Task complexity.}
We first assess the overall difficulty of the two new KB retrieval tasks by considering absolute model performances. Across both datasets, MAP and NDCG scores range around 10 to 20 percent, scoring on the lower side compared to similar retrieval datasets~\cite{thakur2021beir}, underscoring the complexity of the tasks. Performance on the PO dataset is consistently 6 percentage points (pp) lower than on PTM across all models, including MAP, NDCG, and Entity Recall, except in the case of EDEL, suggesting that the PO task is the more challenging of the two.

\textbf{Zero-shot baseline performances.} 
In the zero-shot setting, BM25 performs competitively with neural models on PO, and outperforms them on PTM. This aligns with prior findings that BM25 generalizes well even in domain transfer scenarios~\cite{thakur2021beir}. In contrast, MedCPT struggles out-of-the-box on these specialized biomedical retrieval tasks.

\textbf{Comparing EDEL to zero-shot competitors.}
Not surprisingly, our fine-tuned EDEL models achieves far superior results on both datasets across all metrics when compared to the zero-shot baselines. On the PO dataset, EDEL achieves an NDCG@10 score of 18.62 percent outperforming the baselines by more than 7 pp. Similar improvements of around 7 pp are observed for NDCG@50 (23.66 percent in total) and of around 5 pp for MAP@10 (14.07 in total) and for MAP@50 (15.24 percent in total). EDEL is also better in identifying documents containing potential relevant answer entities as measured by the EntityRecall@10 and @50 metrics. On the PO dataset, EDEL achieves recall scores of 53.51 and 73.48 percent respectively outperforming the next best baselines by 20 pp. These trends also hold true in the PTM dataset albeit less pronounced. EDEL outperforms the best zero-shot baseline by more than 3 pp across the NDCG and MAP metrics and more than 5 pp in Entity Recall.

\textbf{Comparing EDEL to the state-of-the-art model MedCPT.}
We now focus on comparing EDEL in an in-domain setting with the MedCPT variant further fine-tuned on the KB datasets. Fine-tuning MedcPT improves its performance by more than 3 pp in NDCG and MAP scores on the PO dataset and by more than 5 pp on the PTM dataset compared to its pre-trained variant. On the PO dataset, the fine-tuned MedCPT variant is able to outperform all zero-shot baselines but still trails BM25 on the PTM dataset. Overall, EDEL consistently surpasses the fine-tuned MedCPT model, gaining more than 3 pp across all the NDCG, MAP and Entity Recall metrics across both datasets, demonstrating the effectiveness of its tailored training regimen, including hard negatives and the layered, contrastive loss.

\begin{table*}[htbp]
  \centering
  \caption{Ablation studies by leaving out individual components of EDEL. Numbers of the best performing model in each column are marked in bold, the second best are underlined.} 
    \begin{tabular}{l|rr|rr|rr|rr|rr|rr}
    \toprule
    \textbf{Dataset} ($\rightarrow$) & \multicolumn{6}{c|}{\textbf{Precision Oncology}} & \multicolumn{6}{c}{\textbf{Post-translational Modifications}}  \\
    & \multicolumn{2}{c|}{NDCG} & \multicolumn{2}{c|}{MAP} & \multicolumn{2}{c|}{Entity Recall} & \multicolumn{2}{c|}{NDCG} & \multicolumn{2}{c|}{MAP} & \multicolumn{2}{c}{Entity Recall} \\
    \textbf{Model Name} ($\downarrow$) & \multicolumn{1}{c}{@10} & \multicolumn{1}{c|}{@50} & \multicolumn{1}{c}{@10} & \multicolumn{1}{c|}{@50} & \multicolumn{1}{c}{@10} & \multicolumn{1}{c|}{@50} & \multicolumn{1}{c}{@10} & \multicolumn{1}{c|}{@50} & \multicolumn{1}{c}{@10} & \multicolumn{1}{c|}{@50} & \multicolumn{1}{c}{@10} & \multicolumn{1}{c}{@50}  \\ \midrule
    EDEL (-MultiMargin,-NoisyPos, & & & & & & & & & & & & \\
    \hspace{25pt}-HardNeg)  & 14.41 & 16.83 & 12.48 & 13.01 & 23.11 & 45.45 & 8.43 & 11.69 & 5.95 & 6.65 & 28.67 & 43.19 \\
    EDEL (-MultiMargin,-NoisyPos)  & 16.28 & 17.09 & 13.00 & 13.20 & 30.30 & 40.91 & \textbf{24.49} & \textbf{27.24} & \textbf{20.47} & \textbf{21.12} & 39.96 & 51.61 \\
    EDEL (-MultiMargin)  & \underline{17.77} & \underline{21.36} & \textbf{14.65} & \textbf{15.53} & \underline{40.91} & \underline{61.36} & 22.64 & 25.59 & 18.60 & 19.30 & \underline{43.73} & \underline{57.53} \\ \midrule
    EDEL  & \textbf{18.62} & \textbf{23.66} & \underline{14.07} & \underline{15.24} & \textbf{53.41} & \textbf{73.48} & \underline{24.24} & \underline{26.47} & \underline{19.73} & \underline{20.21} & \textbf{50.72} & \textbf{61.65} \\
    \bottomrule
    \end{tabular}%
  \label{tab:20250226_edel_ablations}%
\end{table*}%

\subsection{Ablation studies}
Table~\ref{tab:20250226_edel_ablations} shows results from an ablation study where we systematically disable more and more components of the EDEL model.

The full EDEL model performs generally the best or second-best across all evaluation metrics. While its MAP and NDCG scores are generally within 1 pp of the best ablation variant, it outperforms all ablations by a wide margin in Entity Recall, by roughly 12 pp on PO and 4–11 pp on PTM. This highlights the key contribution of the layered contrastive loss, which appears particularly helpful at identifying documents containing relevant entities in their abstracts (see the full model compared to the EDEL (-MultiMargin) variant). This makes sense as NDCG and MAP score calculation can not distinguish between noisy, gold documents one one side and complete positives on the other in our setting. But to achieve a high Entity Recall, a model is incentivized to rank those noisy examples lower. 

Removing the layered loss and excluding noisy positives in EDEL (-MultiMargin, -NoisyPos) is neutral concerning MAP/NDCG scores (positive on PTM and negative on PO) but causes a further substantial drop in Entity Recall. This suggests that the addition of noisy positives can improve model performance but works better when combined with the MultiMargin loss function.

In the full ablation (-MultiMargin, -NoisyPos, -HardNeg), the model is trained like a standard dense retriever~\cite{karpukhin_dense_2020, jin_medcpt_2023}, using only a small set of complete positives and randomly sampled negatives. The resulting model performs the worst out of all EDEL ablations, comparably to the fine-tuned MedCPT on PO yet falling behind again on PTM. The worse performance on PTM may be due to the lack of preceding retrieval training in the ablation model, whereas MedCPT might benefit more from its prior fine-tuning on 255 M query-document pairs.


Comparing the full ablation (-MultiMargin,-NoisyPos,-HardNeg) to the model that retains hard negatives, EDEL (-MultiMargin, \mbox{-NoisyPos)}, allows us to isolate the contribution of our hard negative sampling strategy. Including structured, informative negatives from the KB improves performance across all metrics, confirming prior findings~\cite{xiong_approximate_2020, hofstatter_efficiently_2021}. This supports our use of KB-derived negatives as an effective strategy for biomedical retrieval tasks.

\begin{table*}[ht]
  \centering
  \caption{Sample queries and their retrieved documents. Key entities are marked in italics. The first two samples show instances of successful retrieval, the latter two of non-optimal retrieval.}
  \resizebox{0.8\textwidth}{!}{%
    \begin{tabular}{cccc}
    \toprule
    \textbf{Query} & \textbf{Doc Retrieved} & \textbf{Top-k} &  \textbf{Doc Curated?} \\
    \midrule
    \makecell[l]{1.\label{tab:discussion_examples_1} Treatment for gene FGFR1 and\\ variant ZNF198::FGFR1?} & \makecell{17698633: ... The viability of Ba/F3 cells \\ transformed to IL3 independence by \textit{ZNF198-FGFR1} or \\ BCR-FGFR1 was specifically inhibited by \textit{TKI258}...} & 2  & No \\
    \makecell[l]{2.\label{tab:discussion_examples_2} Treatment for gene PTPRT and\\ promoter hypermethylation?} & \makecell{31316618: ... We further show that \textit{PTPRT} \\ \textit{promoter methylation} is significantly associated with ...\\ responsiveness to \textit{STAT3 inhibitors} in clinical development ...} & 11   & Yes \\ \midrule
    \makecell[l]{3.\label{tab:discussion_examples_3} Treatment for gene KDR and\\ variant A1065T?} & \makecell{19723655: \textit{KDR activating mutations} \\in human angiosarcomas are sensitive \\ to specific \textit{kinase inhibitors} ...} & >50    & Yes \\
    \makecell[l]{4.\label{tab:discussion_examples_4} Treatment for gene CTAG2 and\\ variant overexpression?} & \makecell{34998699: [Irrelevant Context] ... Upregulation of STAG2 \\ and NCOR1 and down regulation of ARID1A \\ and UTX genes and their targeting miRNAs were \\ associated with UC non-response to BCG ...} & 1    & No \\
    \bottomrule
    \end{tabular}
  }
  \label{tab:discussion_examples}
\end{table*}

\section{Discussion}
\label{sec:discussion}

We now discuss two key aspects of our approach in more detail:
(i) the sensitivity of EDEL to specific values chosen for the margin hyperparameters, and  
(ii) qualitative insights into the behavior of the model through case studies.

\textbf{Impact of the choice for margin value hyperparameters.}
Table~\ref{tab:20250227_margin_value_sensitivity} evaluates how the addition of specific negative samples as well as the choice of their margin value $\mu$ affect model performance. We augment the EDEL (-MultiMargin) ablation with additional hard negatives of class 2 (these negatives closely resemble noisy positives, see Appendix Tables~\ref{tab:margin_classes_negative_po}, \ref{tab:margin_classes_ptm_negative}) and test three different values as margin values $\mu$ for them:  
(i) $\mu = 0.8$ (same as other negatives), (ii) $\mu = 0.0$ (same as positives), and (iii) $\mu = 0.4$ (in between).

Evaluated on the PTM dev set, setting $\mu = 0.8$ interestingly even improves MAP and NDCG but reduces Entity Recall, likely underestimating other noisy but relevant documents. In contrast, setting $\mu = 0.4$ and $\mu = 0.0$ degrades performance across all metrics, likely due to contradictory supervision signal provided to the model. This shows that small changes to the choice of negatives to sample from and their assigned margin values can substantially impact retrieval quality and must be tuned carefully.


\begin{table}[htbp]
  \centering
  \caption{Ablation studies by adding negative samples with different margin values on the PTM development set.}
    \resizebox{\linewidth}{!}{%
    \begin{tabular}{l|rr|rr|rr}
    \toprule
    \textbf{Dataset} ($\rightarrow$) & \multicolumn{6}{c}{\textbf{Post-translational Modifications}}  \\
    Development set & \multicolumn{2}{c|}{NDCG} & \multicolumn{2}{c|}{MAP} & \multicolumn{2}{c}{Entity Recall} \\
    \textbf{Model Name} ($\downarrow$) & \multicolumn{1}{c}{@10} & \multicolumn{1}{c|}{@50} & \multicolumn{1}{c}{@10} & \multicolumn{1}{c|}{@50} & \multicolumn{1}{c}{@10} & \multicolumn{1}{c}{@50}  \\ \midrule
    BM25  & 17.12 & 19.88 & 12.47 & 13.12 & 42.07 & 50.34 \\ \midrule
    EDEL (-MultiMargin)  & 22.74 & 25.57 & 18.88 & 19.44 & 41.03 & 52.76 \\ 
    \hspace{2pt} + Class 2 Neg with $\mu=0.8$ & 26.05 & 28.73 & 22.41 & 22.93 & 39.66 & 51.03 \\
    \hspace{2pt} + Class 2 Neg with $\mu=0.4$ & 20.74 & 22.90 & 17.21 & 17.69 & 42.07 & 52.41 \\
    \hspace{2pt} + Class 2 Neg with $\mu=0.0$ & 20.17 & 21.90 & 16.82 & 17.14 & 39.66 & 51.38 \\
    \bottomrule
    \end{tabular}%
    }
  \label{tab:20250227_margin_value_sensitivity}%
\end{table}%

\textbf{Case studies.}
Table~\ref{tab:discussion_examples} shows examples of retrieved documents by EDEL.
Example 1 demonstrates a relevant document retrieved by EDEL that is not annotated in the gold KB, suggesting that standard evaluation metrics like MAP and NDCG may underestimate model performance.  
Example 2 highlights a case where EDEL correctly assigns high relevance to a curated KB document.

Example 3 shows a classic miss of a retrieval system. A relevant document is ranked too low, potentially due to missing mention of the variant in the abstract. Example 4 shows a typical mistake for dense retrievers as they do not match keywords directly. The retrieved document provides no relevant context and does not mention the query gene at all.

\section{Conclusion}
\label{sec:conclusion}

We introduced the KB curation retrieval task, which focuses on identifying evidence documents for extending arbitrary $n$-ary relations to $(n+1)$-ary structures. To support this task, we presented two new biomedical retrieval datasets, Precision Oncology and Post-Translational Modification, covering complex, real-world relation schemes.

To address the challenges of this task, we proposed EDEL, a dense retriever that leverages the structure of KB records to sample informative, hard negatives and applies a layered contrastive loss function to model varying degrees of document relevance. Our experiments demonstrate that EDEL outperforms both zero-shot and fine-tuned baselines, highlighting the effectiveness of combining structured KB signals with tailored retrieval objectives.

\begin{acks}
We thank Ninon De Mecquenem, Samuele Garda and Oğuz Serbetci for reading and revising the manuscript. Xing David Wang is supported by the DFG as part of the research unit CompCancer (No. RTG2424).
\end{acks}


\bibliographystyle{ACM-Reference-Format}
\bibliography{edel-references}

\appendix

\section{Definition of margin classes}
\label{app:ssec:margin_classes}

\begin{table}[htbp]
\setlength\tabcolsep{4pt} 
\caption{Margin classes for the PO dataset. We report their margin values $\mu$ and their evaluation under the KB matching function $f$ or text matching function $g$, respectively. "-" indicates that the entity type is neither evaluated under $f$ nor $g$.}
  \begin{subtable}[t]{\linewidth}
    \centering
    \caption{Positive sample classes. The KB matching function $f$ evaluates to 1 for all entity types, i.e., $f(\text{Gene})$ = $f(\text{Variant}) = f(\text{Treatment}) = 1$.}
      \begin{tabular}{rr|rrr|r}
        \toprule
        & & \multicolumn{3}{c|}{Eval of Text Matching $g(x)$} & \\
        Class & $\mu $ & Gene & Variant & Treatment & Samples \\
        \midrule
        1     & 0     & 1 & 1 & 1 &                  827  \\
        2     & 0.2   & 1 & 0 & 1 &              2,065  \\
              &       & 0 & 1 & 1 &              \\
        3     & 0.6   & 1 & 1 & 0 &                  218  \\
        4     & 0.8     & - & - & - & 0          \\
        5     & 1.0   & 1 & 0 & 0 &              1,117  \\
              &       & 0 & 1 & 0 &              \\
              &       & 0 & 0 & 1 &              \\
        6     & 1.2   & 0 & 0 & 0 &                  200  \\
        \bottomrule
      \end{tabular}%
      \label{tab:margin_classes_positive_po}
  \end{subtable}
  \\
  \begin{subtable}[t]{\linewidth}
    \centering
    \caption{Negative sample classes. The text matching function $g$ evaluates to 1 for all entity types, i.e., $g(\text{Gene})$ = $g(\text{Variant}) = g(\text{Treatment}) = 1$. Exception is the class 2, where $g(\text{Variant})=0$ indicated by the asterisk *.}
      \begin{tabular}{rr|rrr|r}
        \toprule
        & & \multicolumn{3}{c|}{Eval of KB Matching $f(x)$} & \\
        Class & $\mu$  & Gene & Variant & Treatment & Samples \\
        \midrule
        1     & 0.0     & - & - & - & 0            \\
        2     & 0.2   & 1  & 0*    & 1 &              16,775 \\
        3     & 0.6   & 1  & 0     & 1 &              9,912 \\
              &       & 1  & 1     & 0 &                    \\
        4     & 0.8   & 1  & 0     & 0 &              149,083  \\
              &       & 0  & 1     & 1 &             \\
              &       & 0  & 0     & 1 &            \\
              &       & \multicolumn{3}{c|}{BM25 negatives, same gene} &             \\
        5     & 1.0   & 0  & 1     & 0 &            176,079  \\
              &       & 0  & 0     & 0 &            \\
              &       & \multicolumn{3}{c|}{Random PubMed negatives} &              \\
        6     & 1.2     & - & - & - & 0            \\
        \bottomrule
      \end{tabular}%
      \label{tab:margin_classes_negative_po}
  \end{subtable}
  \label{tab:margin_classes_po}
\end{table}

The KB records in the PO dataset consist of the three entities Gene, Variant, and Treatment, each linked to a document reference (PubMed ID). Each record forms query-document associations where Gene and Variant comprise the query, and PubMed IDs serve as document references. The PTM KB similarly includes four entities: Protein, PTM Type, Residue, and Catalyst.

To move beyond binary relevance labels (positive or negative), we assign a fine-grained margin value $\mu$ to each query-document pair. These values are used in the layered contrastive loss (Eq.~\ref{eq:loss}) to capture partial relevance for noisy positives or differing degrees of irrelevance among negatives. A value of $\mu = 0$ indicates high relevance; values between 0 and 1 represent partial relevance; values above 1 indicate irrelevance (up to a max of 2).

To manage model complexity, we define a small number of margin classes, each corresponding to a fixed value of $\mu$. All samples within a class share the same margin. Assignment to a margin class is based on evaluations of two functions across entity types:
(i) The KB matching function $f$, which compares entities between KB records and
(ii) the text matching function $g$, which checks whether an entity or its synonyms are mentioned in the abstract.

Samples with more entities matching under $f$ are considered closer to a given positive and hence harder negatives. Samples with fewer entities matching under $g$ are considered noisier and less reliable as positives.

We assign samples not derived from KB records to a shared margin class, e.g., BM25 or random PubMed negatives. Tables~\ref{tab:margin_classes_po} and~\ref{tab:margin_classes_ptm} summarize all margin classes for PO and PTM, including function evaluations, assigned margin values, and sample counts.

After tuning on the dev set, we retain four additional non-binary margin classes (besides the standard positive/negative classes). This strikes a balance between expressiveness and model complexity.

\subsection{Positive sample classes}
\label{app:ssec:positives}

Tables~\ref{tab:margin_classes_positive_po} and~\ref{tab:margin_classes_ptm_positive} show the complete list of positive (i.e., complete and noisy positive) classes for PO and PTM.
For positives, the KB matching function $f$ must evaluate to 1 for all entity types, otherwise the example is a negative. In PO, the three entity types yield $2^3=8$ unique text matching patterns for the text matching function $g$, and in PTM with four types, we get $2^4=16$ possible combinations.
Margin assignment follows these principles:
\begin{itemize}
\item \textbf{More matched entities lead to a lower margin:} 
   For instance, a sample with $g(\text{Gene})=1$ and $g(\text{Treatment})=1$, $g(\text{Variant})=0$ is less noisy (and thus gets a smaller margin) than one where $g(\text{Gene})=1$, $g(\text{Treatment})=0$ and $g\text{(Variant})=0$.
\item \textbf{Specific entity types may be more important:} 
   In PO, Treatment entities as the target entities are prioritized. For instance, the positive Class 2 with $g(\text{Treatment})=1$ is ranked more relevant than Class 3 where $\text{g(Treatment})=0$, even if both match two entities.
\item The remaining margins are chosen via \textbf{grid search} and \textbf{domain-informed heuristics} (see Section~\ref{sec:discussion}). While this yielded a strong configuration, the search space was not fully exhausted.
\end{itemize}

\subsection{Negative sample classes}
\label{app:ssec:negatives}

Negative sample construction mirrors the logic for positives. Here, the text matching function $g$ is set to 1 for all entity types to minimize label noise, except for PO Class 2, where $g(\text{Variant})=0$ to introduce particularly difficult negatives, and some samples from PTM classes 4 and 5.
In addition to KB-derived negatives, we include BM25 negatives (same Gene/Protein) and random PubMed samples. Tables~\ref{tab:margin_classes_negative_po} and~\ref{tab:margin_classes_ptm_negative} provide a full overview of all margin classes and their corresponding values.

We assume that any KB-linked document not referenced in a positive record is a true negative. While this may introduce noise due to incomplete curation, we verified that this risk is limited: About 1.8 percent of negative abstracts contain the same variant, and about 7.8\% mention the same gene or its synonyms. Thus, fewer than 10\% of negatives are potentially mislabeled.

\section{Model training}
\label{app:sec:training}

We conducted training on one NVIDIA A100 GPU with 80 GB GPU memory with a batch size of $n=32$ and epoch number of eight. We use AdamW~\cite{DBLP:conf/iclr/LoshchilovH19} with a maximum learning rate of 2e-5 and linearly decaying learning rate schedule after an initial warm-up phase.
Maximum input token length into the model is 512, longer inputs being truncated. Title and abstract of each document are separated by the special token [SEP].
After model training, we pre-compute the embeddings for all PubMed abstracts and store them in a Faiss FlatIPIndex~\cite{johnson_billion-scale_2021}.
Code and datasets will be made available online upon publication.

\begin{table}[htbp]
\setlength\tabcolsep{4pt} 
\caption{Margin classes for the PTM dataset. We report their margin values $\mu$ and their evaluation under the KB matching function $f$ or text matching function $g$, respectively. "-" indicates that the entity type is neither evaluated under $f$ nor $g$.}
  \begin{subtable}[t]{\linewidth}
    \centering
    \caption{Positive sample classes. The KB matching function $f$ evaluates to 1 for all entity types, i.e., $f(\text{Protein})$ = $f(\text{PTM type}) = f(\text{Residue}) = f(\text{Catalyst}) = 1$.}
    \begin{tabular}{rr|rrrr|r}
    \toprule
    & & \multicolumn{4}{c|}{Eval of Text Matching $g(x)$} & \\
    Class & $\mu$ & Protein & PTM & Residue  & Catalyst & Samples \\
    \midrule
    1     & 0     & 1 & 1 & 1 & 1 &                  563  \\
    2     & 0.4  & 1 & 1 & 0 & 1 &              1,179  \\
      & & 1 & 0 & 1 & 1 &                \\
      & & 0 & 1 & 1 & 1 &                \\
    3     & 0.6   & 1 & 1 & 1 & 0 &                  1,405  \\
    & & 1 & 1 & 0 & 0 &                \\
    & & 1 & 0 & 1 & 0 &                \\
    & & 1 & 0 & 0 & 1 &                \\
    & & 0 & 1 & 1 & 0 &                \\
    & & 0 & 1 & 0 & 1 &                \\
    & & 0 & 0 & 1 & 1 &                \\
    4     & 0.8   & 1 & 0 & 0 & 0 &                  1,294  \\
    & & 0 & 1 & 0 & 0 &                \\
    & & 0 & 0 & 1 & 0 &                \\
    & & 0 & 0 & 0 & 1 &                \\
    5     & 1.0   & 0 & 0 & 0 & 0 &                  51  \\
    6     & 1.2   & - & - & - & - &                  0  \\
    \bottomrule
    \end{tabular}%
    \label{tab:margin_classes_ptm_positive}
  \end{subtable}
  \\
  \begin{subtable}[t]{\linewidth}
    \centering
    \caption{Negative sample classes. The text matching function $g$ evaluates to 1 for all entity types, i.e., $g(\text{Protein})$ = $g(\text{PTM type}) = g(\text{Residue}) = g(\text{Catalyst}) = 1$. Exceptions are classes 4 and 5, where $g(\text{Protein})=0$ indicated by the asterisk *.}
    \begin{tabular}{rr|rrrr|r}
    \toprule
    & & \multicolumn{4}{c|}{Eval of KB Matching $f(x)$} & \\
    Class & $\mu$ & Protein & PTM & Residue & Catalysts &  Support  \\
    \midrule
    1     & 0     & - & - & - & - &                  0  \\
    2     & 0.4   & 1 & 1 & 0 & 1 &                  2,015  \\
    3     & 0.6   & 1 & 0 & 0 & 1 &                  1,260  \\
    & & 1 & 1 & 0 & 0 &             \\
    4     & 0.8   & 1 & 0 & 0 & 0 &                  45,736  \\
    & & 1*& - & - & 1 &            \\
    & & 0 & 0 & 0 & 1 &           \\
    & & \multicolumn{4}{c|}{BM25 negatives, same gene} &   \\
    5 & 1.0 & 1*& - & - & 0 & 43,473           \\
    & & 0 & 0 & 0 & 0 &            \\
    6 & 1.2 & \multicolumn{4}{c|}{Random PubMed negatives} &  37,980  \\
    \bottomrule
    \end{tabular}%
    \label{tab:margin_classes_ptm_negative}
  \end{subtable}
  \label{tab:margin_classes_ptm}
\end{table}










\end{document}